\documentclass[5p,10pt]{elsarticle}
\biboptions{numbers,sort&compress}
\usepackage{newtxtext,newtxmath}
\usepackage{amsmath}
\usepackage{graphicx}
\usepackage{booktabs}
\usepackage{multirow}
\usepackage{array}
\usepackage{float}
\usepackage[colorlinks,citecolor=blue,urlcolor=blue,linkcolor=blue]{hyperref}
\usepackage{microtype}
\usepackage{ragged2e}
\usepackage{caption}
\journal{Physics Letters B}
\newcommand{\dd}{\mathrm{d}}
\newcommand{\be}{\begin{equation}}
	\newcommand{\ee}{\end{equation}}

\newcommand{\beq}{\begin{eqnarray}}
	\newcommand{\eeq}{\end{eqnarray}}

\begin{document}

\begin{frontmatter}

\title{Confronting Inflation and Reheating with Observations: Improved Predictions}

 \author[label1,label2]{Ying-Ying Ye}
 \author[label1,label2]{Bao-Min Gu\corref{cor1}}
 \ead{gubm@ncu.edu.cn}
 \cortext[cor1]{Corresponding author.}
 \address[label1]{Department of physics, Nanchang University, Nanchang 330031, China}
 \address[label2]{Center for Relativistic Astrophysics and High Energy Physics, Nanchang University, Nanchang 330031, China}

\begin{abstract}
Using the latest observational data, we constrain the inflationary dynamics and the subsequent reheating epoch. Predictions for both phases can be significantly improved by employing numerically computed results compared to the slow-roll approximations. These results enable a more accurate reassessment of the observational viability of inflationary models, provide tighter constraints on the reheating history, and help lift the degeneracies in the predictions of inflation and reheating dynamics. Given current observational bounds, this enables a more accurate understanding of the early universe physics.
\end{abstract}


\end{frontmatter}


\section{Introduction}
\label{section.1}

Cosmic inflation is the leading framework addressing the horizon and flatness problems of standard Big Bang cosmology \cite{Starobinsky1980,Guth1981,Linde1982,Olive1990,Linde:2014nna} and providing the quantum origin of the seeds for the large-scale structure formation \cite{Mukhanov1981,Guth1982,Sasaki1986,Copeland:1997et,Sriramkumar2009}. Key inflationary observables, such as the scalar spectral index $n_{s}$ and tensor-to-scalar ratio $r$, are tightly constrained by measurements of the cosmic microwave background (CMB) observations \cite{Planck:2018jri,BICEP:2021xfz,ACT:2025fju,ACT:2025tim}. Following inflation, the universe undergoes a reheating phase, during which the inflaton decays, transferring energy to Standard Model particles and reheating the universe—thereby initiating the hot Big Bang era \cite{Starobinsky1982,Dolgov:1982th,Albrecht1982,Abbott1982,Kofman1994,Traschen:1990sw,Kofman:1997yn}. Precise characterization of this reheating epoch is thus essential for connecting inflationary predictions to the universe’s subsequent evolution.

Studies of inflation and reheating typically rely on the slow-roll approximation, which assumes that the inflaton’s dynamics are dominated by its potential, neglecting the kinetic contributions \cite{Mukhanov:1990me,Baumann:2009ds}. However, near the end of inflation, this assumption breaks down as the potential no longer dominates \cite{Baumann:2009ds,Auclair:2024udj}, causing the slow-roll approximation to fail and bias the predictions of inflation and reheating. Hence, a numerical solution of the inflaton’s equations of motion is necessary, providing a more precise evolution for the inflaton dynamics \cite{Nandi:2024dil}.

Recent cosmological constraints from the combined dataset-incorporating ACT DR6 \cite{ACT:2025fju,ACT:2025tim}, Planck 2018
\cite{Planck:2018vyg,Planck:2018jri}, BICEP/Keck 2018 \cite{BICEP:2021xfz} and DESI data \cite{DESI:2024uvr,DESI:2024mwx} (P-ACT-LB-BK18)-yield a scalar spectral index of $n_{s} = 0.9743 \pm 0.0034$. The new results place renewed pressure on the viability of existing inflationary scenarios, challenging their ability to remain consistent with data \cite{Kallosh:2025rni,Aoki:2025wld,Berera:2025vsu,Dioguardi:2025vci,Salvio:2025izr,Dioguardi:2025mpp,Gao:2025onc,
	Drees:2025ngb,Haque:2025uri,Zharov:2025evb,Haque:2025uis,Yogesh:2025wak,Addazi:2025qra,Yi:2025dms,Kallosh:2025ijd,
	Peng:2025bws,Saini:2025jlc,Haque:2025uga,Heidarian:2025drk,Gao:2025viy,Pal:2025ewf,Mohammadi:2025gbu,Zharov:2025zjg,
	SidikRisdianto:2025qvk,Lynker:2025wyc,Peng:2025vda}.
This also motivates the adoption of refined theoretical frameworks to reassess these models and explore novel scenarios \cite{Adhikari:2019uaw,Khan:2022odn,Yogesh:2024zwi,Yogesh:2024mpa,Zahoor:2025nuq,Kouniatalis:2025orn,Haque:2025uri,Maity:2025czp,Liu:2025qca,Saini:2025jlc,Yi:2025dms,
	Yin:2025rrs,Han:2025cwk,Heidarian:2025drk,Chakraborty:2025jof,Pallis:2025gii,Kim:2025dyi,
	Gao:2025viy,Chakraborty:2025jof,Wolf:2025ecy,Okada:2025lpl,Gialamas:2025kef,Gialamas:2025ofz}. Moreover, analyses based solely on the slow-roll approximation often suffer from parameter degeneracies. For example, within $\alpha$-attractor models, the spectral index $n_{s}$ shows weak dependence on specific parameters \cite{Mishra2021}, and distinct inflationary models can produce nearly degenerate reheating histories \cite{Cabella2017}.

In this work we use the numerical approach, solving the full equations of motion for the inflaton field to obtain more accurate predictions of inflationary dynamics and of the reheating history. Our analysis focuses on two representative classes of models: the well-studied $\alpha$-attractors and the power-law potentials. This treatment yields significantly improved predictions for inflation and reheating. In contrast to the slow-roll approximation, the numerical approach
(i) determines the duration of inflation with high precision;
(ii) incorporates higher-order corrections to the spectral index, leading to accurate estimates of inflationary observable; and
(iii) captures the full inflaton dynamics, particularly during the critical end phase of inflation. 
The numerical approach offers a decisive advantage in breaking degeneracies between inflationary models. This is evident in two key aspects. First, within the $\alpha$-attractor framework 
(i) under the slow-roll approximation, the spectral index $n_{s}$ is largely insensitive to the parameter $\alpha$ at fixed power-law exponent $n$;
(ii) conversely, for fixed $\alpha$, both $n_{s}$ and the tensor-to-scalar ratio $r$ exhibit weak dependence on $n$.
The numerical analysis resolves this insensitivity. 
Second, the numerical method distinguishes reheating histories that are degenerate under the slow-roll approximation. For example, it differentiates between $\alpha$-attractor E-models and Fibre Inflation models, which otherwise yield nearly identical reheating predictions in the analytical framework.

The structure of this paper is as follows. In Sec.~\ref{section.21}, we revisit the observable of inflation and establish the connection between inflation and reheating based on the cosmic expansion history. Sec.~\ref{section.22} presents the framework for predicting inflation and reheating dynamics using the slow-roll approximation. In Sec.~\ref{section.23}, we highlight the limitations of the slow-roll approach, introduce the numerical methodology, and apply it to obtain refined predictions for inflation and reheating. Sec.~\ref{section.3} demonstrates how the numerical approach resolves the degeneracies of inflation and reheating, using the results of Sec.~\ref{section.2}, and derives tighter constraints on the inflaton decay rate $\Gamma$ in perturbative reheating. We conclude and outline future directions in Sec.~\ref{section.4}. In this paper we adopt the metric signature $(-,+,+,+)$ and set the reduced Planck mass to unity, $M_{\rm Pl}^2 = 1/8\pi G = 1$. Dot and prime represents the derivative with respect to the cosmic time $t$ and the e-folds $N$, respectively.

\section{The Prediction Of Inflation And Reheating}
\label{section.2}
\subsection{General formalism}
\label{section.21}
Let us consider a spatially flat Friedmann–Lemaître–Robertson–Walker (FLRW) background with line element
\begin{equation}
	\dd s^{2} = -\dd t^{2} + a^{2}(t)\dd\vec{x}^{2}. 
	\label{eq.1}
\end{equation}
During inflation, the energy density is dominated by a scalar field—the inflaton—whose dynamics govern  
the background expansion. The Klein–Gordon equation for the inflaton field and the Friedmann equation are
\beq
\ddot{\phi} + 3H\dot{\phi} + V_{,\phi} &=& 0,\\
\frac{1}{2}\dot{\phi}^{2} + V(\phi) &=& 3H^{2}.
\label{eq.2,3}
\eeq
In the slow-roll regime of inflation, the kinetic energy of the inflaton is subdominant, $\frac{1}{2}\dot{\phi}^2 \ll V(\phi)$, and its acceleration is negligible compared to the Hubble friction term, $|\ddot{\phi}| \ll |H\dot{\phi}|$. These conditions imply smallness of the Hubble slow-roll parameters,
\begin{equation}
	\epsilon_{1} = -\frac{\dot{H}}{H^{2}} = \frac{\dot{\phi}^{2}}{2 H^{2}} \ll 1, \quad
	\epsilon_{2} = \frac{\dot{\epsilon}_{1}}{H \epsilon_{1}} \ll 1.
	\label{eq.4}
\end{equation}
Equivalently, in terms of the potential slow-roll parameters, the conditions become
\begin{equation}
	\epsilon_V \equiv \frac{1}{2} \left( \frac{V_\phi}{V} \right)^2 \ll 1, \quad
	\eta_V \equiv \frac{|V_{\phi\phi}|}{V} \ll 1.
	\label{eq.5}
\end{equation}
Under these conditions, the background undergoes an approximately quasi–de Sitter expansion. To leading order in slow-roll parameters, the scalar spectral index and tensor-to-scalar ratio are given by
\begin{figure}[htb]
	\centering
	\includegraphics{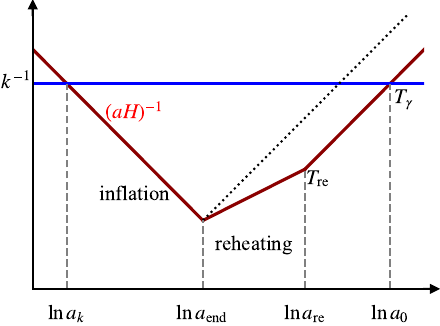}
	\caption{Evolution of the comoving Hubble radius from $a_k$ to $a_0$, illustrating the phases of inflation, reheating, and standard expansion. The dotted line corresponds to instantaneous reheating with an effective equation-of-state parameter $w_{\phi} = 1/3$. The post-inflationary evolution of the universe influences the present-day horizon scale, necessitating that viable reheating histories be consistent with current observations~\cite{Liddle2003}.}
	\label{figure.1}
\end{figure}
\begin{equation} 
	n_s = 1 - 6\epsilon_V(\phi_*) + 2\eta_V(\phi_*), \quad
	r = 16\epsilon_V(\phi_*),
	\label{eq.6}
\end{equation}
where $\phi_*$ denotes the field value at horizon crossing. These observables are tightly constrained by CMB measure, particularly by the recent ACT DR6 data.

After inflation, the universe undergoes a reheating phase, during which the energy stored in the inflaton field is transferred to Standard Model particles, repopulating the thermal bath and initiating the radiation-dominated era that precedes Big Bang Nucleosynthesis. It was pointed out that the predictions of reheating can be constrained by that of inflation \cite{Liddle2003,Dai2014,Cook2015,Maiti:2024nhv}. In the following we revisit this procedure.

To connect inflation with the subsequent reheating phase, let us consider the evolution of the comoving Hubble radius from the horizon exit of the pivot scale $k$ to the present epoch \cite{Liddle2003}, as illustrated in Fig.~\ref{figure.1}. The total expansion from horizon exit to present can be decomposed as
\begin{equation}
	\ln\left(\frac{a_0}{a_k}\right) = \ln\left(\frac{a_0}{a_{\mathrm{re}}}\right) + \ln\left(\frac{a_{\mathrm{re}}}{a_{\mathrm{end}}}\right) + \ln\left(\frac{a_{\mathrm{end}}}{a_k}\right).
	\label{eq.7}
\end{equation}
It is convenient to define 
\begin{eqnarray}
	N_k &= \ln\left(\frac{a_{\mathrm{end}}}{a_k}\right), \label{eq:(5)} \\
	N_{\mathrm{re}} &= \ln\left(\frac{a_{\mathrm{re}}}{a_{\mathrm{end}}}\right),
	\label{eq.8,9}
\end{eqnarray}
corresponding to the duration from horizon exit to the end of inflation, and from the end of inflation to the end of reheating, respectively.
From the end of reheating to the present epoch, the conservation of entropy implies
\begin{equation}
	\dd(g_{*s} a^3 T^3) = 0,
	\label{eq.10}
\end{equation}
leading to the relation
\begin{equation}
	g_{*s,\mathrm{re}} a_{\mathrm{re}}^3 T_{\mathrm{re}}^3 = g_{*s,0} a_0^3 T_\gamma^3.
	\label{eq.11}
\end{equation}
Here, $g_{*s}$ denotes the effective number of relativistic degrees of freedom associated with entropy, with the subscripts ``re" and ``0" referring to the end of reheating and the present time, respectively.
At the present epoch, the effective number of relativistic degrees of freedom contributing to entropy is given by
\begin{equation}
	g_{*s,0} = g_\gamma + g_\nu \left( \frac{T_\nu}{T_\gamma} \right)^3,
	\label{eq.12}
\end{equation}
where $T_\gamma$ and $T_\nu$ are the background temperatures of the photons and neutrinos at present-day, respectively. Typically, $g_\gamma = 2$, $g_\nu = 21/4$, and $T_\nu^3 = (4/11) T_\gamma^3$.
Combining with Eq~.(\ref{eq.11}) and Eq~.(\ref{eq.12}), the scale factor at the end of reheating relative to today is given by
\begin{equation}
	\frac{a_{\mathrm{re}}}{a_0} = \left( \frac{43}{11 g_{\mathrm{*s,re}}} \right)^{1/3} \frac{T_\gamma}{T_{\mathrm{re}}}.
	\label{eq.13}
\end{equation}

During reheating, the inflaton undergoes coherent oscillations about the minimum of its potential, which can be approximated by a monomial form $V(\phi) \propto \phi^n$. In this regime, the field behaves as an effective fluid characterized by an equation-of-state parameter $w_\phi$. Assuming the Hubble expansion time scale is much longer than the oscillation period, $w_\phi$ can be obtained by averaging the pressure and the energy density over one oscillation cycle, yielding \cite{Ford:1986sy,Garcia:2020wiy}
\begin{equation}
	w_\phi = \frac{\langle p_\phi \rangle}{\langle \rho_\phi \rangle} \simeq \frac{n - 2}{n + 2},
	\label{eq.w}
\end{equation}
where $\langle \cdots \rangle$ denotes a cycle average. Once the form of the potential is specified, $w_\phi$ is uniquely determined. For $\alpha$-attractors Eq.~(\ref{eq.28}), $n=2$ corresponds to $\omega_\phi=1/3$. Hence, models with $n<2$ have soft equation of state while $n>2$ corresponds to stiff equation of state. Treating $w_\phi$ as constant, the energy densities at the end of inflation and at the end of reheating obey the relation
\begin{equation}
	\rho_{\mathrm{re}} = \left( \frac{a_{\mathrm{re}}}{a_{\mathrm{end}}} \right)^{-3(1 + w_{\phi})} \rho_{\mathrm{end}}.
	\label{eq.14}
\end{equation}
At the end of reheating, the energy density is also given by the thermal relation
\begin{equation}
	\rho_{\mathrm{re}} = \frac{\pi^2}{30} g_{\mathrm{re}} T_{\mathrm{re}}^4,
	\label{eq.15}
\end{equation}
where the $g_{\mathrm{re}}$ is the effective number of relativistic degrees of freedom. At the end of reheating, all relativistic species are nearly in thermal equilibrium \cite{Shtanov:1994ce,Lozanov2019}, allowing one to identify $g_{*s} = g_{\mathrm{re}}$.
Eqs.~(\ref{eq.14}) and (\ref{eq.15}) then give an expression for the reheating temperature
\begin{equation}
	\ln T_{\mathrm{re}} = -\frac{3}{4}(1 + w_{\phi}) N_{\mathrm{re}} + \frac{1}{4} \ln \rho_{\mathrm{end}} - \frac{1}{4} \ln\left( \frac{\pi^2}{30} g_{\mathrm{re}} \right).
	\label{eq.16}
\end{equation}
For the pivot scale that exits the horizon during inflation and re-enters today, it satisfies
\be
k = a_{k} H_{k} = a_{0}H_{0}.
\label{eq.17}
\ee
Using Eqs.~(\ref{eq.7}), (\ref{eq.13}), (\ref{eq.16}) and (\ref{eq.17}), we obtain the relation between inflation and reheating
\be
N_{\text{re}} = \frac{4}{3w_{\phi}-1} \left(N_{k} + \frac{1}{4} \ln{3 H^{2}_{e}} - \ln{H_{k}} - 61.65\right),
\label{eq.18}
\ee
\be
T_{\text{re}} = \left(\frac{90 H^{2}_{e}}{\pi^{2} g_{\text{re}}}\right)^{\frac{1}{4}} e^{-\frac{3}{4} (1+w_{\phi})N_{\text{re}}},
\label{eq.19}
\ee
where we adopt $k=0.05~\mathrm{Mpc}^{-1}$, $g_{\mathrm{re}} = 106.75$, and $T_\gamma = 2.725~\mathrm{K}$. Given a model-dependent relation between the scalar spectral index $n_s$, tensor-to-scalar ratio $r$, and the quantities $N_k$, $H_e$, and $H_k$, these expressions enable predictions for reheating in terms of inflationary observables.
\subsection{Predictions in slow-roll}
\label{section.22}
In this subsection, we present the observable of inflation and reheating under the slow-roll approximation. The background dynamics are governed by
\begin{align}
	3 H \dot{\phi} + V_{,\phi} &\simeq 0, \label{eq.20} \\
	H^{2} &\simeq \frac{1}{3} V(\phi). \label{eq.21}
\end{align}
The number of $e$-folds before the end of inflation is
\beq
&&N^{\text{sr}}_{k} = -\int_{\phi^{\text{sr}}_{*}}^{\phi^{\text{sr}}_{e}} \frac{V}{V_{,\phi}} \dd \phi
=-\int_{\phi^{\text{sr}}_{*}}^{\phi^{\text{sr}}_{e}} \frac{\dd \phi}{\sqrt{2\epsilon_V}} .
\label{eq.22}
\eeq
The Hubble parameter at horizon crossing and at the end of inflation are given by
\beq
&&H_{k} \simeq H^{\text{sr}}_{k} = \pi \sqrt{rA_{s}/2},
\label{eq.23}
\\
&&H_{e} \simeq H^{\text{sr}}_{e} = \sqrt{V(\phi^{\text{sr}}_{e})/3},
\label{eq.24}
\eeq
where $\phi^{\text{sr}}_{e}$ is determined by $\epsilon_{V}(\phi) \simeq 1$ and $A_{s}$ is the amplitude of the scalar power spectrum evaluated at $k=0.05~\text{Mpc}^{-1}$. In slow-roll approximation, the spectral index $n_{s}$ and the tensor-to-scalar ratio $r$ are given by
\beq
&&n_{s} = 1 - 6\epsilon_{V}(\phi^{\text{sr}}_*) + 2 \eta_{V}(\phi^{\text{sr}}_*),
\label{eq.25}
\\
&&r = 16\epsilon_{V}(\phi^{\text{sr}}_*).
\label{eq.26}
\eeq
\begin{figure*}[!t] 
	\centering
	\begin{minipage}[t]{0.28\textwidth} %
		\centering
		\includegraphics[width=\linewidth]{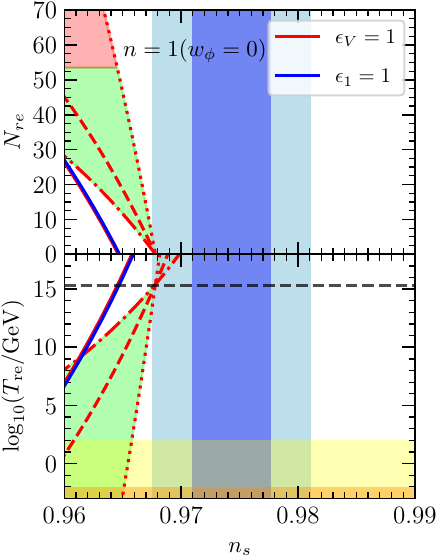} %
	\end{minipage}
	\hfill %
	\begin{minipage}[t]{0.28\textwidth} %
		\centering
		\includegraphics[width=\linewidth]{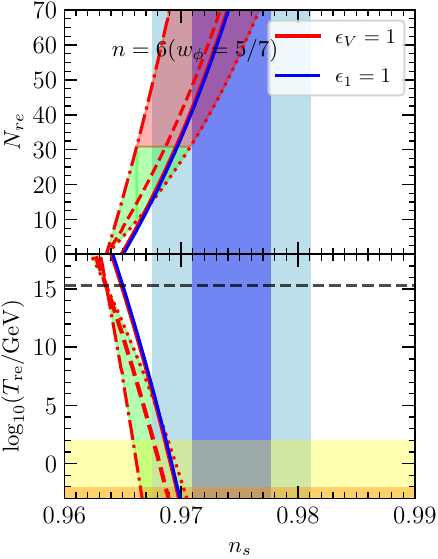} %
	\end{minipage}
	\hfill
	\begin{minipage}[t]{0.28\textwidth} %
		\centering
		\includegraphics[width=\linewidth]{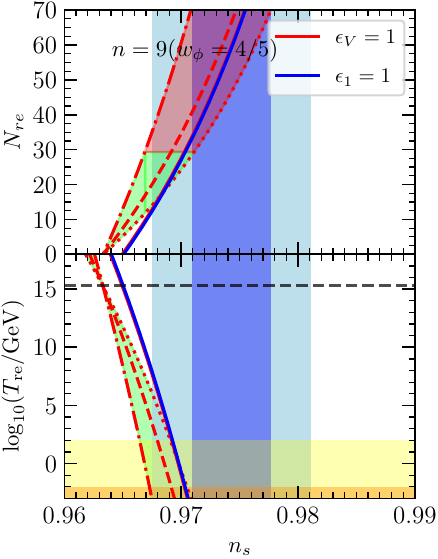} %
	\end{minipage}
	\begin{minipage}[t]{0.28\textwidth} %
		\centering
		\includegraphics[width=\linewidth]{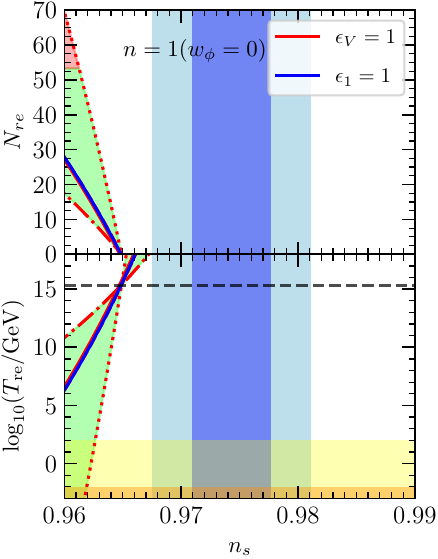} %
	\end{minipage}
	\hfill %
	\begin{minipage}[t]{0.28\textwidth} %
		\centering
		\includegraphics[width=\linewidth]{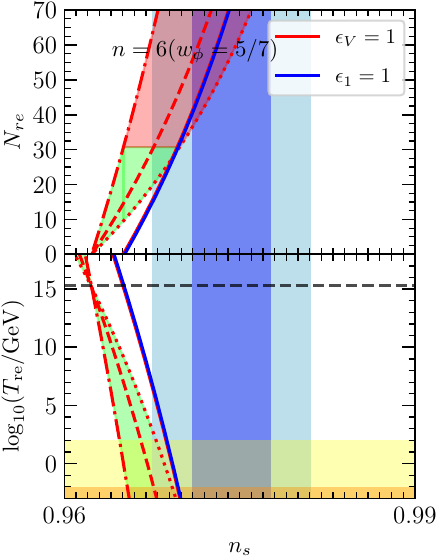} %
	\end{minipage}
	\hfill
	\begin{minipage}[t]{0.28\textwidth} %
		\centering
		\includegraphics[width=\linewidth]{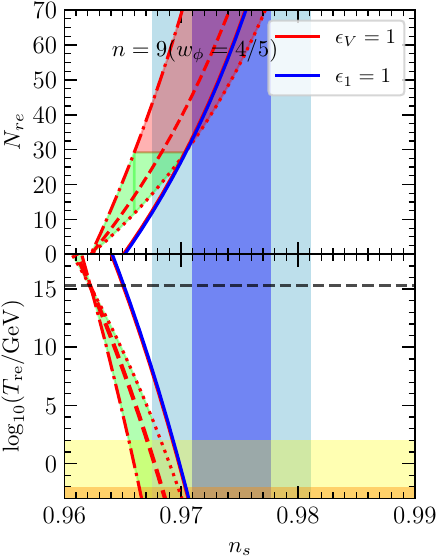} %
	\end{minipage}
	\begin{minipage}[t]{0.28\textwidth} %
		\centering
		\includegraphics[width=\linewidth]{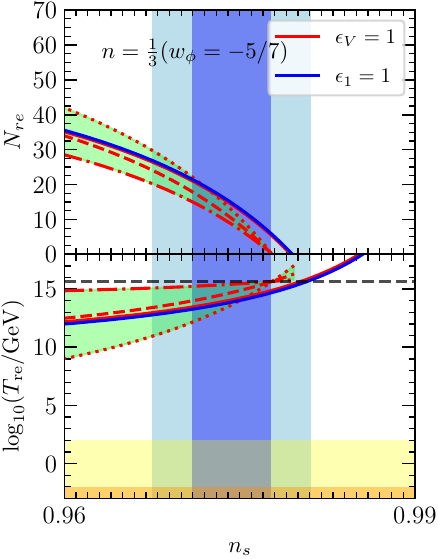} %
	\end{minipage}
	\hfill %
	\begin{minipage}[t]{0.28\textwidth} %
		\centering
		\includegraphics[width=\linewidth]{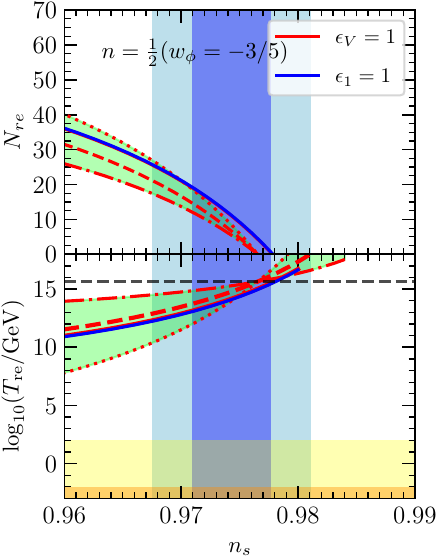} %
	\end{minipage}
	\hfill
	\begin{minipage}[t]{0.28\textwidth} %
		\centering
		\includegraphics[width=\linewidth]{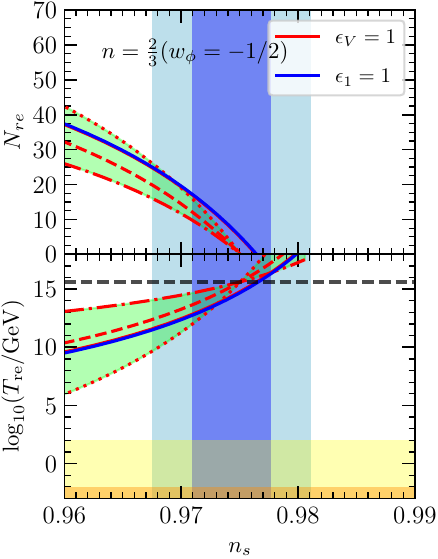} %
	\end{minipage}
	\caption{Predictions for reheating from the E-model ($\alpha=10$), T-model ($\alpha=10$), and power-law inflation model (top to bottom) are shown, where within each panel the solid line denotes the slow-roll approximation, with red and blue curves corresponding to inflation endpoints defined by $\epsilon_V=1$ and $\epsilon_1=1$, respectively, and dashed lines represent the numerical results. The dark blue band is the 1$\sigma$ constraint on $n_s$ from the P-ACT-LB-BK18 data, while the light blue band shows the 2$\sigma$ constraint. The black dashed line denotes the upper bound on $T_{\mathrm{re}}$, corresponding to instantaneous reheating with $N_{\mathrm{re}}=0$. The dark yellow and light yellow regions correspond to the lower bounds of $T_{\mathrm{re}}$ from BBN \cite{Hasegawa:2019jsa,Hannestad:2004px,DeBernardis:2008zz,deSalas:2015glj} and the electroweak (EW) scale, respectively. The region shaded in red are the parameter space of $N_{\mathrm{re}}$ that excluded by Big Bang nucleosynthesis (BBN) requirements by using Eq.~(\ref{eq.19}).}
	\label{figure.2}
\end{figure*}

The quantities designated by superscript ``sr" are computed under slow-roll approximation.
Exploiting the relation
\be
H^{\text{sr}}_{e} = \sqrt{V(\phi^{\text{sr}}_{e})/3} = \sqrt{\frac{V(\phi^{\text{sr}}_{e})}{V(\phi_{*}^{\text{sr}})}} H^{\text{sr}}_{k},
\label{eq.27}
\ee
once $\phi^{\text{sr}}_{e}$ is determined, Eqs.~(\ref{eq.22}), Eqs.~(\ref{eq.23}) and Eqs.~(\ref{eq.24}) can be expressed as functions of the spectral index $n_{s}$ via the intermediate field value $\phi_{*}^{\text{sr}}$. Subsequently, using Eqs.~(\ref{eq.18}) and Eqs.~(\ref{eq.19}), the observable $n_{s}$ yields predictions for the reheating history, namely $N_{\mathrm{re}}(n_{s})$ and $T_{\mathrm{re}}(n_{s})$. 
To investigate the implications for reheating, let us consider several representative inflationary models given by
\be
V(\phi) = 
\begin{cases}
	\Lambda^{4} \left(1 - e^{-\sqrt{\frac{2}{3\alpha}} \phi}\right)^{2n}, &\text{E-model},\\
	\Lambda^{4} \left(\tanh \left(\frac{\phi}{\sqrt{6\alpha}}\right)\right)^{2n}, &\text{T-model},\\
	V_{0} \phi^{n},
	&\text{power-law}.
\end{cases}
\label{eq.28}
\ee
In the limit of a large number of e-folds $N_{k}$, the slow-roll expressions for the scalar spectral index and tensor-to-scalar ratio are given by~\cite{Kallosh2013,Roest2014}
\be
n_{s} \simeq 
\begin{cases}
	1 - \frac{2}{N^{\text{sr}}_{k}}, &\text{E, T-model},\\
	
	1 - \frac{n+2}{2N^{\text{sr}}_{k}}, &\text{power-law},
\end{cases}
\label{eq.29}
\ee
and 
\be
r \simeq
\begin{cases}
	\frac{12 \alpha}{(N^{\text{sr}}_{k})^{2}}, &\text{E, T-model},\\
	
	\frac{4n}{N^{\text{sr}}_{k}}, &\text{power-law}.
\end{cases}
\label{eq.30}
\ee
These relations enable predictions for the reheating phase within the slow-roll approximation. We consider the three representative models and present the resulting reheating predictions in Fig.~\ref{figure.2} with solid lines. In particular, we focus on the equation of state parameter obtained by Eq.~(\ref{eq.w}) for each model. The results are consistent with those presented in previous literatures.
\subsection{Predictions in numerical approach}
\label{section.23}
The slow-roll approximation presumes potential dominance during inflation. However, as inflation ends, the inflaton’s kinetic energy grows comparable to the potential, rendering the slow-roll treatment invalid and potentially introducing inaccuracies in the full dynamical evolution. 
As a result, the slow-roll framework fails to accurately determine the end of inflation or resolve the full inflaton dynamics, introducing uncertainties into the predictions of inflation and reheating. Moreover, the scalar spectral index $n_s$ is commonly expressed using potential slow-roll parameters,
\beq
\epsilon_{V} &=& \frac{1}{2} \left(\frac{V_{,\phi}}{V}\right)^{2},\\
\eta_{V} &=& \left|\frac{V_{,\phi \phi}}{V}\right|.
\label{eq.31,32}
\eeq
To improve upon this, we employ the Hubble flow parameters, defined recursively as \cite{Liddle2000}
\be
\epsilon_{i+1} = \frac{\dd\ln{\epsilon_{i}}}{\dd N}, \quad i \geq 1,
\label{eq.33}
\ee
where $N$ is the number of e-folds during inflation and $\epsilon_1=-H'/H$. To overcome the limitations of the slow-roll approximation, we solve the full equation of motion for the inflaton field numerically,
\be
\phi{''}  +\left(3 - \frac{1}{2} \phi{'}^{2}\right) \left(\phi'+\frac{V_{,\phi}}{V}\right) = 0,
\label{eq.34}
\ee
with suitable initial conditions $\phi(N_0)$ and $\phi'(N_0)$. Applying the inverse function yields $N(\phi)$.
The Hubble parameter is then obtained from
\be
H^{\text{num}} =\sqrt{\frac{2V}{6-\phi{'}^{2}}}.
\label{eq.35}
\ee
It also allows one to get $N(H)$.
To further enhance the precision, we compute the next-to-leading-order expressions for the spectral index and the tensor-to-scalar ratio \cite{Stewart1993, Martin2024},
\begin{eqnarray}
	n_{s} &=& 1 - 2\epsilon_{1}(N) - \epsilon_{2}(N) -2(\epsilon_{1}(N)^{2}) -  \nonumber \\ &&(3+2C_{1})\epsilon_{1}(N)\epsilon_{2}(N) - C_{1}\epsilon_{2}(N)\epsilon_{3}(N), \label{eq.36}  \\
	r &=&16\epsilon_{1}(N)(1+C_{1}\epsilon_{2}(N)),
	\label{eq.37}
\end{eqnarray}
where $C_1 =  \ln 2 + \gamma-2  \simeq -0.7296$, and $\gamma$ is the Euler–Mascheroni constant. Solving Eq.~(\ref{eq.34}) yields the full inflationary dynamics, including $H(N)$ and $\epsilon_i(N)$, thereby enabling the derivation of the reheating parameters $N_{\mathrm{re}}$ and $T_{\mathrm{re}}$ as a function of $n_s$. This accurately captures the inflationary dynamics beyond the slow-roll approximation and yields more reliable predictions of inflation and reheating process. 

Following the end of inflation, the inflaton field oscillates near the minimum of its potential, marking the onset of the reheating phase. During this phase, the inflaton's behavior is modeled as a fluid with an equation-of-state parameter $w_{\phi}$. As previously noted, treating the inflaton oscillations as periodic and assuming the cosmic expansion timescale significantly exceeds the oscillation period, the cycle-averaged energy density and pressure yield the ideal equation of state given by Eq.~(\ref{eq.w}). However, strictly speaking, deviations from Eq.~(\ref{eq.w}) may arise throughout reheating. Consequently, for numerical purposes, we empirically select equation-of-state parameters within a range around the value given by Eq.~(\ref{eq.w}), i.e., $w_{\phi}\pm1/5$. 

\begin{figure*}[t]
	\centering
	\begin{minipage}[t]{0.43\textwidth}
		\centering
		\includegraphics[width=7.5cm]{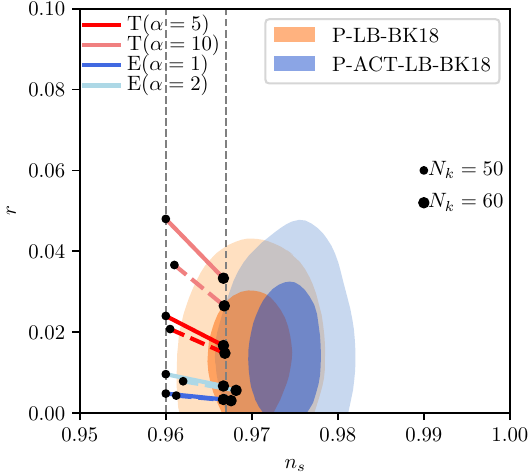}
	\end{minipage}
	\hfill
	\begin{minipage}[t]{0.43\textwidth}
		\centering
		\includegraphics[width=7.5cm]{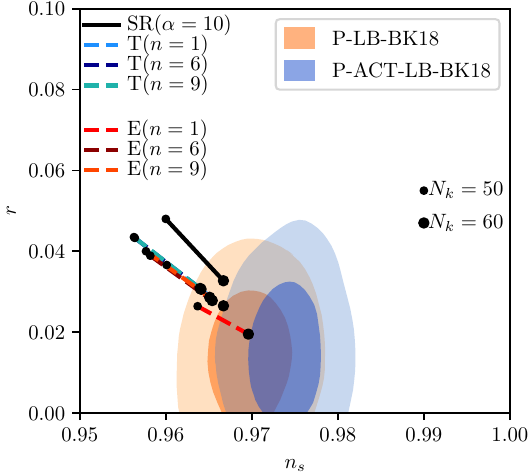}
	\end{minipage}
	\caption{Predictions for the scalar spectral index $n_s$ and tensor-to-scalar ratio $r$ in $\alpha$-attractor inflation models, comparing the slow-roll approximation (solid lines) and numerical results (dashed lines). The left panel shows E- and T-models with $n=1$ and varying $\alpha$, while the right panel shows E,T-models with $\alpha=10$ for varying $n$. The shaded regions denote the 2$\sigma$ (light blue) and 1$\sigma$ (dark blue) confidence contours from the combined P-ACT-LB-BK18 dataset.}
	\label{figure.3}
\end{figure*}

Fig.~\ref{figure.3} shows the predictions of inflation with numerical approach and slow-roll one. We see that the spectral index and the tensor-to-scalar ratio can be predicted more accurately. These results allow us to reexamine inflationary models under latest observational constraints. For example, the predictions of the E-model with $n=1$ and $\alpha=10$ in slow-roll lie outside the allowed $2\sigma$ region and may be excluded, while in numerical approach we see this model is consistent with latest observations.

\begin{table*}[ht!]
	\centering
	\begin{tabular*}{\textwidth}{@{\extracolsep{\fill}}l c c c c c@{}}
		\hline
		Model & \( n \) & \( N_{\text{re}}^{\text{sr}} \) & \( N_{\text{re}}^{\text{num}} \) & \( \log_{10} T_{\text{re}}^{\text{sr}} / \text{GeV} \) & \( \log_{10} T_{\text{re}}^{\text{num}} / \text{GeV} \) \\
		\hline
		E-model & 1 & Not allowed & [0, 2.21] & Not allowed & [14.72, 15.43] \\
		& 6 & [15.97, 171.80] & [23.19, 30.86] & [-2, 6.37] & [-2, 2.29] \\
		& 9 & [12.85, 140.05] & [20.20, 29.34] & [-2, 7.73] & [-2, 3.36] \\
		\hline
		T-model & 1 & Not allowed & Not allowed & Not allowed & Not allowed \\
		& 6 & [15.28, 171.11] & [30.00, 30.71] & [-2, 6.67] & [-2, -1.60] \\
		& 9 & [12.41, 139.61] & [24.70, 29.23] & [-2, 7.92] & [-2, 5.67] \\
		\hline
		power-law & 1/3 & [0, 25.85] & [0, 22.19] & [13.25, 15.66] & [13.60, 15.66] \\
		& 1/2 & [0, 25.63] & [0, 23.31] & [12.31, 15.65] & [12.61, 15.65] \\
		& 2/3 & [0, 26.39] & [0, 23.06] & [11.34, 15.63] & [11.87, 15.63] \\			
		\hline
	\end{tabular*}
	\caption{Reheating predictions for the $\alpha$-attractors ($\alpha$ = 10) and power-law inflationary models within the 2$\sigma$ confidence interval for $n_s$ from the P-ACT-LB-BK18 dataset.}
	\label{table.1}
\end{table*}
\begin{table*}[ht!]
	\centering
	\begin{tabular*}{\textwidth}{@{\extracolsep{\fill}}l c c c c c c c c c@{}}
		\hline
		& & \multicolumn{2}{c}{\(\phi_{e}\)} & \multicolumn{3}{c}{\(N_{k}\)} & \multicolumn{3}{c}{\(H_{e}\)} \\
		\cline{3-4} \cline{5-7} \cline{8-10}
		Model & \(n\) & \(\phi_{e}^{\text{sr}}\) & \(\phi_{e}^{\text{num}}\) & \(N_{k}^{\epsilon_{V}}\) & \(N_{k}^{\epsilon_{1}}\) & \(N_{k}^{\text{num}}\) & \(H_{e}^{\epsilon_V}\) & \(H_{e}^{\epsilon_1}\) & \(H_{e}^{\text{num}}\) \\
		\hline
		E-model 
		& 1 & 1.21 & 0.83 & 50.55 & 50.77 & 51.91 & $4.82\times10^{-6}$ & $3.48\times10^{-6}$ & \(4.27\times10^{-6}\) \\
		& 6 & 4.49 & 3.99 & 67.11 & 67.43 & 68.83 & $2.05\times10^{-6}$ & $1.19\times10^{-6}$ & \(1.45\times10^{-6}\) \\
		& 9 & 5.64 & 5.12 & 67.87 & 68.20 & 69.32 & $1.80\times10^{-6}$ & $1.03\times10^{-6}$ & \(1.27\times10^{-6}\) \\
		\hline
		T-model 
		& 1 & 1.39 & 0.98 & 51.28 & 51.52 & 52.57 & $3.49\times10^{-6}$ & $2.71\times10^{-6}$ & \(3.31\times10^{-6}\) \\
		& 6 & 5.91 & 5.38 & 69.47 & 69.82 & 71.01 & $1.43\times10^{-6}$ & $7.98\times10^{-7}$ & \(9.78\times10^{-7}\) \\
		& 9 & 7.38 & 6.84 & 69.57 & 69.93 & 71.11 & $1.34\times10^{-6}$ & $7.59\times10^{-7}$ & \(9.30\times10^{-7}\) \\
		\hline
		power-law 
		& 1/3 & 0.24 & 0.06 & 43.82 & 43.89 & 47.80 & $1.05\times10^{-5}$ & $9.87\times10^{-6}$ & \(1.21\times10^{-5}\) \\ 
		& 1/2 & 0.35 & 0.13 & 47.13 & 47.24 & 49.79 & $9.97\times10^{-6}$ & $9.29\times10^{-6}$ & \(1.14\times10^{-5}\) \\
		& 2/3 & 0.47 & 0.21 & 50.39 & 50.52 & 52.49 & $9.33\times10^{-6}$ & $8.37\times10^{-6}$ & \(1.03\times10^{-5}\) \\
		\hline
	\end{tabular*}
	\caption{Comparison of inflationary quantities for the $\alpha$-attractor ($\alpha=10$) and power-law models. Entries show slow-roll results—with superscripts $\epsilon_V$ and $\epsilon_1$ denoting end of inflation defined by $\epsilon_V=1$ and $\epsilon_1=1$, respectively—and full numerical results (superscript $\mathrm{num}$). All cases use the same $\phi_*$, $\phi_{\mathrm{e}}$ is taken as the value satisfying $\epsilon_V=1$ or $\epsilon_1=1$ as indicated.}
	\label{table.2}
\end{table*}

\begin{figure*}[t]
	\centering
	\begin{minipage}[t]{0.45\textwidth}
		\centering
		\includegraphics[width=6.0cm]{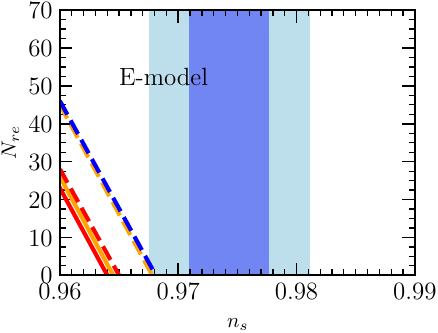}
	\end{minipage}
	\hfill
	\begin{minipage}[t]{0.45\textwidth}
		\centering
		\includegraphics[width=6.0cm]{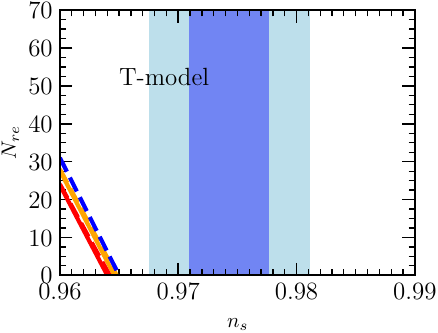}
	\end{minipage}
	\caption{Predictions for reheating in $\alpha$-attractor inflation models at $n=1$ ($w_{\phi}=0$), as $\alpha$ varies. The left panel shows results for the E-model, while the right panel shows those for the T-model. In both figures, solid curves represent results under the slow-roll approximation, and dashed curves denote numerical results. The values $\alpha=1$, $10$, and $100$ correspond to red, orange, and blue, respectively.}
	\label{figure.4}
\end{figure*}

We also compare the numerical predictions of reheating with those obtained under the slow-roll approximation in Fig.~\ref{figure.2} and Table~\ref{table.1},
Fig.~\ref{figure.2} reveals two important results. First, we see that within the 2$\sigma$ confidence level of  $n_{\mathrm{s}}$,
the numerical method imposes more accurate constraints on the duration of reheating and the reheating temperature. For instance, in the E-model with $n=6$, the slow-roll approximation yields a reheating duration of 15.97$\sim$171.80 e-folds. In the numerical approach, however, combined with BBN constraints, the duration $N_{\text{re}}$ is restricted in the range 23.19$\sim$30.86 e-folds (see Table \ref{table.1} for detailed values). Second, since the P-ACT-LB-BK18 data gives a larger value of spectral index compared to the previous results, the reheating history of inflation models are severely constrained by the new results. 
The larger value of the scalar spectral index $n_s$ suggested by recent observations disfavors reheating histories with $w_{\phi}<1/3$ in the T-model and in the $\alpha\lesssim10$ limit of the E-model. For $\alpha\gtrsim10$, the predicted reheating history of the E-model lies in the edge of the $2\sigma$ observational bound. 
As an example, Fig.~\ref{figure.4} illustrates the reheating duration for $w_{\phi} = 0$ constrained by P-ACT-LB-BK18. 

The prediction of reheating from inflation depends critically on the scalar spectral index, $n_{\mathrm{s}}$. We see that the results from numerical methods deviate from those of slow-roll approximation significantly. To quantify the cumulative error in the dynamical evolution, we compare the two approaches starting from the same initial field value. We now analyze the origin of these differences.

Starting from the same initial field value, the numerical and slow-roll methods yield different e-folding numbers $N_{k}$. Through their respective $N_{k}-n_{s}$ relations, this leads to different spectral indices $n_{s}$ and thus different reheating histories. The discrepancy arises from:
\begin{enumerate}
	\item \textbf{The e-folds during inflation $N_{k}$}
	The numerical approach solves the full Klein-Gordon equation and determines the actual inflation endpoint, resulting in a different integration interval, while the slow-roll method uses an analytic formula.
	\item \textbf{Hubble parameter at the end of inflation $H_{e}$}
	The numerical result includes both kinetic and potential energy, whereas the slow-roll estimate considers only the potential energy 
	\item \textbf{The relation between $N_{k}$ and $n_{s}$} The numerical method determines $n_{s}$ from $N_{k}$ by expanding to second order in the Hubble slow-roll parameters Eq.~(\ref{eq.36}), whereas the slow-roll approximation uses only the first-order expansion in the potential slow-roll parameters Eq.~(\ref{eq.29}). Hence, for the same $N_{k}$, the numerical result provides a more precise value of $n_{s}$, which underlies the discrepancy in the predicted inflationary observables.
\end{enumerate}

During the early stage of inflation the numerical solution and the slow-roll approximation are in close agreement; consequently, for a common initial condition the Hubble parameter at horizon exit, $H_k$, is essentially identical. From the inflation–reheating relation Eq.~(\ref{eq.18}), the remaining quantities that can affect reheating are the e-folding number $N_k$ and the Hubble parameter at the end of inflation $H_e$; the entries in Table~\ref{table.2} indicate that the effect of $N_k$ is dominant.

When the endpoint of inflation is aligned—e.g., defined uniformly by $\epsilon_1 = 1$ ($\phi_e^{\mathrm{num}}$)—the predictions for inflation and reheating from the numerical and slow-roll approaches still differ. This indicates that the remaining discrepancy originates purely from corrections to the inflationary dynamics. In practice, changing the inflation endpoint alters the integration interval for computing $N_k$ (e.g., from $\phi_e^{\mathrm{SR}}$ to $\phi_e^{\mathrm{num}}$) as well as the value of $H_e$, which, according to Eqs. (\ref{eq.18}) and (\ref{eq.19}), also affect $N_{\mathrm{re}}$ and $T_{\mathrm{re}}$. However, this effect is small. For comparison, we present in Table~\ref{table.2} the differences between the numerical and slow-roll results with the same endpoints, as well as those with different endpoints.

Furthermore, varying the inflation endpoint does not alter the $N_k$–$n_s$ relation in the large-$N_k$ limit. Consequently, in the $N_{\mathrm{re}}$–$n_s$ ($T_{\mathrm{re}}$–$n_s$) plane, such a change influences the curves only through $H_e$, resulting in a negligible shift, as showed in Fig.~\ref{figure.2}. Owing to this robustness of the relation, the choice of endpoint likewise does not affect the predictions for inflation.

\begin{figure}[htb]
	\centering
	\includegraphics[width=6.5cm]{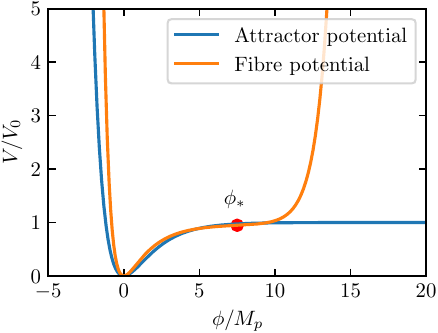}
	\caption{The inflationary potentials for the E-model with $\alpha=2$, $n=1$ (blue) and Fibre Inflation (brown).}
	\label{figure.7}
\end{figure}
\begin{figure*}[t]
	\centering
	\begin{minipage}[t]{0.45\textwidth}
		\centering
		\includegraphics[width=6.0cm]{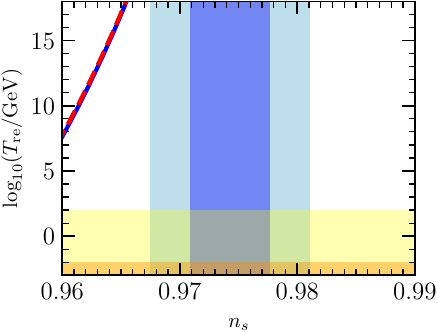}
	\end{minipage}
	\hfill
	\begin{minipage}[t]{0.45\textwidth}
		\centering
		\includegraphics[width=6.0cm]{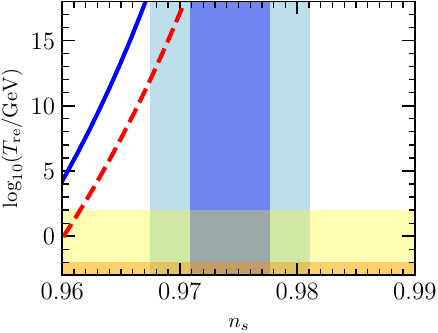}
	\end{minipage}
	\caption{This figure presents predictions for reheating from the E-model (solid) with $\alpha=2$ and $n=1$ and Fibre inflation (dashed). The left panel shows results obtained using the slow-roll approximation, while the right panel displays results from numerical calculations.}
	\label{figure.8}
\end{figure*}

\section{Applications of the numerical results}
\label{section.3}
In this section, we highlight applications of numerical methods, particularly in resolving degeneracies among inflationary models. Such degeneracies arise, for example, when the key observable such as the spectral index 
and the tensor-to-scalar ratio  exhibit weak dependence on the model parameters~\cite{Roest2014,Safaei:2024idb,Bahari:2025ttf}, or when distinct models yield similar reheating predictions. Moreover, the improved precision in determining the reheating temperature through numerical approaches enables tighter constraints on the perturbative decay rate $\Gamma$ of the inflaton.
\subsection{Breaking the degeneracy of the inflation observable}
\label{section.31}
For $\alpha$-attractor models, the predictions for $n_s$ and $r$ from Eqs.(\ref{eq.29}) and (\ref{eq.30}) are consistent with the Planck 2018 observations for typical e-folds $N_k \simeq 50$–$60$, but show mild tension with the combined P-ACT-LB-BK18 data. Another notable limitation is the model’s insensitivity to the potential parameters: (i) for fixed $\alpha$, varying the power $n$ leads to identical predictions for $n_s$ and $r$; (ii) for fixed $n$, $n_s$ remains unchanged under variations in $\alpha$ \cite{Kallosh2013}. Consequently, observationally distinct inflationary scenarios become degenerate at the level of $n_s$ and $r$, posing a challenge for experimental discrimination. As shown in Fig.~\ref{figure.3}, These degeneracies can be alleviated through full numerical analysis. This improvement arises from the more accurate numerical evaluation of $n_s$ and $r$ via Eqs.~(\ref{eq.34}) and Eqs.~(\ref{eq.35}), capturing the inflationary dynamics more precisely. For fixed $\alpha$ and varying $n$, the numerical method yields distinct $(n_s, r)$ predictions, in contrast to the degeneracy observed in the slow-roll case, where models fall along a single curve. 
\subsection{Breaking the degeneracy of reheating predictions}
\label{section.32}
Inflationary models can exhibit degeneracies not only in their predicted observables but also in their reheating dynamics. Specifically, distinct models may yield nearly identical reheating histories, complicating efforts to distinguish between them based on reheating alone. For instance, the E-model with $\alpha = 2$ and $n = 1$, and the string-inspired fiber inflation model \cite{Cicoli2009,Burgess2016} produce similar reheating predictions \cite{Cabella2017}.
\begin{table*}[ht!]
	\centering
	\begin{tabular*}{\textwidth}{@{\extracolsep{\fill}}l c c c c c@{}}
		\hline
		Model & \( n \) & \multicolumn{1}{c}{\( \log_{10} T_{\text{re}}^{\text{sr}} / \text{GeV} \)} & \multicolumn{1}{c}{\( \log_{10} T_{\text{re}}^{\text{num}} / \text{GeV} \)} & \multicolumn{1}{c}{\( \varGamma^{\text{sr}} \)} & \multicolumn{1}{c}{\( \varGamma^{\text{num}} \)} \\
		\hline
		E-model & 1 & Not allowed & [14.72, 15.43] & -- & [$1.60\times10^{-7}$, $4.21\times10^{-6}$] \\
		& 6 & [-2, 6.37] & [-2, 2.29] & [$5.81\times10^{-41}$, $3.50\times10^{-24}$] & [$5.81\times10^{-41}$, $2.21\times10^{-32}$] \\
		& 9 & [-2, 7.73] & [-2, 3.36] & [$5.81\times10^{-41}$, $1.68\times10^{-21}$] & [$5.81\times10^{-41}$, $3.05\times10^{-30}$] \\
		\hline
		T-model & 1 & Not allowed & Not allowed & -- & -- \\
		& 6 & [-2, 6.67] & [-2, -1.6] & [$5.81\times10^{-41}$, $1.27\times10^{-23}$] & [$5.81\times10^{-41}$, $3.66\times10^{-40}$] \\
		& 9 & [-2, 7.92] & [-2, 5.67] & [$5.81\times10^{-41}$, $4.02\times10^{-21}$] & [$5.81\times10^{-41}$, $1.27\times10^{-25}$] \\
		\hline
		power-law & 1/3 & [13.25, 15.66] & [13.60, 15.66] & [$1.84\times10^{-10}$, $1.21\times10^{-5}$] & [$9.21\times10^{-10}$, $1.21\times10^{-5}$] \\
		& 1/2 & [12.31, 15.65] & [12.68, 15.65] & [$2.42\times10^{-12}$, $1.16\times10^{-5}$] & [$1.33\times10^{-11}$, $1.16\times10^{-5}$] \\
		& 2/3 & [11.34, 15.62] & [11.61, 15.62] & [$2.78\times10^{-14}$, $1.01\times10^{-5}$] & [$9.64\times10^{-14}$, $1.01\times10^{-5}$] \\
		\hline
	\end{tabular*}
	\caption{Based on the more precise reheating temperatures obtained from the numerical results, we present further constraints on the inflaton decay rate $\varGamma$.}
	\label{table.3}
\end{table*}
The latter arises from Type IIB string compactifications on K3-fibered Calabi–Yau manifolds \cite{Greene:1996cy}, where Kähler moduli are stabilized via string loop corrections. This mechanism generates a sufficiently flat potential to support inflation. Notably, all tunable parameters enter the potential only through an overall scale $V_0$ and do not affect the slow-roll parameters, which depend solely on the number of inflationary e-folds $N_k$~\cite{Cicoli2009}. This feature imparts a degree of universality to the model’s observational predictions. In what follows, we consider the effective potential of fiber inflation as a case study,
\be
V(\phi) = V_{0} \left(c_{0} + c_{1} e^{-\frac{1}{2} k \phi} + c_{2} e^{-2k \phi} + c_{3} e^{k \phi}\right),
\label{eq.38}
\ee
here $k = 2/\sqrt{3}$, and $c_0$, $c_1$, $c_2$, $c_3$ are numerical coefficients determined by the underlying string compactification \cite{Cicoli2009}. At early times, the potential of fiber inflation admits the approximation
\be
V(\phi) \simeq V_{0} \left(3 - 4 e^{-\frac{\phi}{\sqrt{3}}}\right),
\label{eq.39}
\ee
which closely resembles the early-time behavior of the E-model with $\alpha = 2$, $n = 1$, as shown in Fig.~\ref{figure.7}. Consequently, the spectral index and the number of $e$-folds during inflation satisfy
\be
N_{k} \simeq \frac{2}{1 - n_{s}},
\label{eq.40}
\ee
identical to the relation in Eq.~(\ref{eq.29}), suggesting similar reheating predictions with E-model ($n=1$,$\alpha=2$) under the slow-roll approximation.
However, in numerical approach we see that they exhibit entirely different reheating history, as demonstrated in Fig.~\ref{figure.8}.
\subsection{Constraining the inflaton decay rate $\varGamma$}
\label{section.33}
Based on the more accurate reheating temperatures obtained from the numerical analysis in Sec.~\ref{section.23}, and considering the perturbative reheating process \cite{Abbott1982,Albrecht1982,Dolgov:1982th}, the inflaton decay width and the reheating temperature can be expressed as follows \cite{Lozanov2019} 
\be
T_{re} \simeq \left(\frac{90}{g_{re} \pi^{2}}\right)^{1/4} \sqrt{\varGamma}.
\label{eq.41}
\ee
We derive the improved constraints on the inflaton decay rate $\varGamma$. The enhanced precision of $T_{\mathrm{re}}$ from the numerical approach enables tighter bounds on $\varGamma$, as summarized in Table~\ref{table.3}.
\section{Conclusion}
\label{section.4}
To summarize, we employed numerical methods to investigate inflationary dynamics, using precise inflationary dynamics to predict the inflationary observable and the reheating histories.  Compared to the slow-roll approximation, fully numerical methods provide more accurate predictions for inflation and reheating dynamics. When combined with the latest observational constraints, this allows for a more robust reassessment of inflationary models. We find that certain parameter choices in the $\alpha$-attractor model, previously excluded under the slow-roll approximation, are actually consistent with current data when evaluated numerically. Conversely, it is plausible that some models deemed viable under the slow-roll treatment may be ruled out by numerical analysis, although we do not provide explicit examples in this work. 

The numerical approach also improved the reheating predictions. The improvement primarily stems from the numerical method of calculating $N_{k}$. This enhancement persists even when a unified endpoint for inflation is adopted. This allows for more accurate constraints on the duration and temperature of reheating, leading to a more precise understanding of the reheating history. For example, the perturbative decay rate can be constrained by the reheating temperature.  This, in turn, is crucial for reconstructing the thermal history of the early Universe.

Numerical results help resolve degeneracies in the predictions of inflation and reheating. For example, under the slow-roll approximation, different combinations of $(n, \alpha)$ in $\alpha$-attractor models yield nearly indistinguishable predictions, making them difficult to differentiate. In contrast, numerical computations reveal significant differences between these scenarios. Similarly, while the E-model and Fibre Inflation produce comparable reheating predictions in the slow-roll framework, full numerical analysis shows that their reheating histories are distinguishable. These results highlight the crucial role of numerical methods in accurately probing the physics of the early Universe.

\section{Acknowledgements}
This work is supported by the National Natural Science Foundation of China (Grants No. 12165013) and the Natural Science Foundation of Jiangxi province, China, under Grant No. 20224BAB211026.

\bibliographystyle{elsarticle-num} 
\bibliography{inflation_and_reheating}

\end{document}